\documentclass[12pt]{article}%
\usepackage{mathptmx}
\usepackage{sprinconf}
\usepackage{graphicx}
\usepackage{bm}%
\begin{document}
\title{On a theoretical model for d-wave to mixed s- and d-wave transition in cuprate superconductors}
\author{Y. Krivolapov\footnote{Email: evgkr@tx.technion.ac.il}}
\author{A. Mann}
\author{Joseph L. Birman\footnote{Permanent address: Department of Physics, City College, CUNY, New York, New York 10031}}
\affil{Department of Physics Technion-Israel Institute of
Technology, Haifa 32000 Israel}

\beginabstract
A $U\left(  3\right)$ model proposed by Iachello for
superconductivity in cuprate materials is analyzed. The model
consists of \emph{s} and \emph{d} pairs (approximated as bosons) in
a two-dimensional Fermi system with a surface. The transition occurs
between a phase in which the system is a condensate of one of the
bosons, and a phase which is a mixture of two types of bosons. In
the current work we have investigated the validity of the Bogoliubov
approximation, and we used a reduced Hamiltonian to determine a
phase diagram, the symmetry of the phases and the temperature
dependence of the heat capacity.
\endabstract

For more than a decade there has been a debate regarding the nature
of the symmetry of the macroscopic wavefunction for the copper oxide
superconductors. According to M{\"{u}}ller bulk sensitive
experiments support substantial \emph{s} symmetry, while surface
sensitive experiments yield \emph{d} symmetry for the macroscopic
wave function. M{\"{u}}ller proposed \cite{Muller} that a
reconciliation of the conflicting experiments was possible if the
superconducting wavefunction were a sum of two components, namely,
one of \emph{s} symmetry and one of \emph{d} symmetry, and varied as
a function of the distance from the surface.

Following that work by M{\"{u}}ller, a theoretical framework for cuprate
superconductors was proposed by Iachello \cite{Iachello}, based on the analogy
with atomic nuclei. In the present work we report our analysis of phase
transitions based on Iachello's model. Also we have investigated temperature
dependence of the heat capacity. We propose a macroscopic wavefunction for
realizing space dependence.

In Eq.\ref{ham_iach} we give Iachello's Hamiltonian for $2N$ spin $\frac{1}%
{2}$ fermions on a plane. This is a boson Hamiltonian \cite{OAI} , which
consists of three types of bosons $s$, $d_{+}$ and $d_{-}$.
\begin{eqnarray}\label{ham_iach}%
H &=&\varepsilon_{s}\left(  \hat{s}^{\dag}\hat{s}\right)
+\varepsilon
_{d}\left(  \hat{d}_{+}^{\dag}\hat{d}_{+}+\hat{d}_{-}^{\dag}\hat{d}%
_{-}\right)  +u_{0}\left(  \hat{s}^{\dag}\hat{s}^{\dag}\hat{s}\hat{s}\right)
+\\
 &+& u_{2}\left(  \hat{d}_{+}^{\dag}\hat{d}_{+}^{\dag}\hat{d}_{+}\hat{d}%
_{+}+\hat{d}_{-}^{\dag}\hat{d}_{-}^{\dag}\hat{d}_{-}\hat{d}_{-}\right)
+u_{2}^{\prime}\left(  \hat{d}_{+}^{\dag}\hat{d}_{-}^{\dag}\hat{d}_{+}\hat
{d}_{-}\right)  +\nonumber\\
 &+& v_{0}\left(  \hat{d}_{+}^{\dag}\hat{d}_{+}+\hat{d}_{-}^{\dag}\hat{d}%
_{-}\right)  \left(  \hat{s}^{\dag}\hat{s}\right)  +v_{2}\left(  \hat{d}%
_{+}^{\dag}\hat{d}_{-}^{\dag}\hat{s}\hat{s}+\hat{s}^{\dag}\hat{s}^{\dag}%
\hat{d}_{+}\hat{d}_{-}\right)\nonumber
\end{eqnarray}
Introducing the operators \cite{Iachello}%
\begin{equation}%
\begin{tabular}[c]{lll}
$\hat{n}_{s}=\hat{s}^{\dag}\hat{s}$ & $\hat{n}_{d+}=\hat{d}_{+}^{\dag}\hat
{d}_{+}$ & $\hat{n}_{d-}=\hat{d}_{-}^{\dag}\hat{d}_{-}$\\
$\hat{n}_{d}=\hat{n}_{d+}+\hat{n}_{d-}$ & $\hat{l}=\hat{n}_{d+}-\hat{n}_{d-}$
& $\hat{N}=\hat{n}_{d+}+\hat{n}_{d-}$%
\end{tabular}
\end{equation}
the Hamiltonian can be rewritten as
\begin{eqnarray}
H &=& \varepsilon_{s}\hat{n}_{s}+\varepsilon_{d}\hat{n}_{d}+u_{0}\hat{n}%
_{s}\left(  \hat{n}_{s}-1\right)  +\label{ham_ns_nd}\\
  &+&\frac{1}{2}u_{2}\hat{n}_{d}\left(  \hat{n}_{d}-2\right)  ++\frac{1}%
{2}u_{2}^{\prime}\hat{n}_{d}^{2}+\frac{1}{2}\left(  u_{2}-u_{2}^{\prime
}\right)  \hat{l}^{2}+\nonumber\\
  &+&v_{0}\hat{n}_{d}\hat{n}_{s}+v_{2}\left(  \hat{d}_{+}^{\dag}\hat{d}%
_{-}^{\dag}\hat{s}\hat{s}+\hat{s}^{\dag}\hat{s}^{\dag}\hat{d}_{+}\hat{d}%
_{-}\right) \nonumber
\end{eqnarray}
Its algebraic structure is $U\left(  3\right)  $%
\[%
\begin{tabular}[c]{lll}
$\hat{n}_{s}=\hat{s}^{\dag}\hat{s}$ & $\hat{n}_{d}=\hat{d}_{+}^{\dag}\hat
{d}_{+}+\hat{d}_{-}^{\dag}\hat{d}_{-}$ & $\hat{l}=\hat{d}_{+}^{\dag}\hat
{d}_{+}-\hat{d}_{-}^{\dag}\hat{d}_{-}$\\
$\hat{Q}_{+}=\hat{d}_{+}^{\dag}\hat{d}_{-}$ & $\hat{T}_{-}=-\hat{d}_{-}^{\dag
}\hat{s}+\hat{s}^{\dag}\hat{d}_{+}$ & $\hat{T}_{+}=\hat{d}_{+}^{\dag}\hat
{s}\emph{-}\hat{s}^{\dag}\hat{d}_{-}$\\
$\hat{Q}_{-}=\hat{d}_{-}^{\dag}\hat{d}_{+}$ & $\hat{R}_{+}=\hat{d}_{+}^{\dag
}\hat{s}+\hat{s}^{\dag}\hat{d}_{-}$ & $\hat{R}_{-}=\hat{d}_{-}^{\dag}\hat
{s}+\hat{s}^{\dag}\hat{d}_{+}$%
\end{tabular}
\ \ \ \ \ \ \ \ \
\]
This algebra contains a subalgebra $SO\left(  3\right)  $ formed from the
generators $\hat{T}_{+},\hat{T}_{-}$ , $\hat{l}$ ,\ and a Casimir operator
defined as (note a factor of $2$ difference from Ref.~\cite{Iachello})%
\[
\hat{W}^{2}=\hat{T}_{+}\hat{T}_{-}+\hat{T}_{-}\hat{T}_{+}+\hat{l}^{2}%
\]

Following Iachello this Hamiltonian can be split into an $SO\left(  3\right)
$ symmetric part, which will contain $\hat{l}$ and $\hat{W}^{2}$ terms, and a
$U\left(  1\right)  \otimes U\left(  2\right)  $ symmetric part generated by
$\hat{n}_{s}$ and $\hat{l},\hat{n}_{d},\hat{Q}_{\pm},$ respectively. When the
Hamiltonian has $SO\left(  3\right)  $ symmetry the system is in phase $II$,
which corresponds to a mixture of \emph{s-} and \emph{d-} bosons. When the
Hamiltonian has $U\left(  1\right)  \otimes U\left(  2\right)  $ symmetry the
system is in phase $I$, which corresponds to either an
\emph{s-}boson condensate (phase $Ia)$ or a \emph{d-}boson condensate (phase
$Ib)$\cite{Iachello}.

\begin{figure}[t]
\includegraphics[width=0.9\textwidth]
{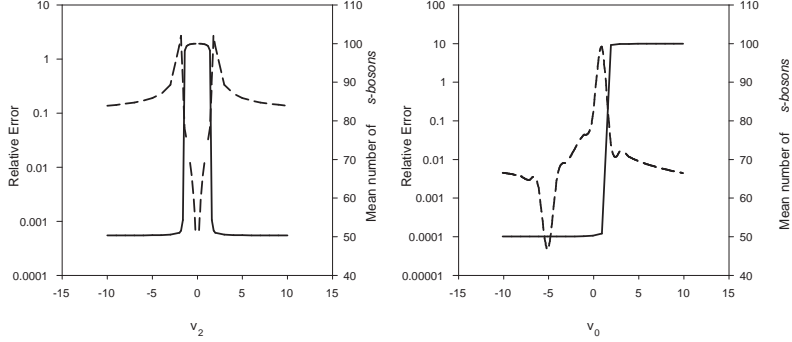}%
\caption{Relative error in the ground state energy (left
\emph{y-}axis, dashed line), calculated using the Bogoliubov
approximation, and exact mean \emph{s-bosons} population (right
\emph{y-}axis, solid line) versus parameters of the Hamiltonian. The
fixed parameters are $N=100,$ $l=0,$
$\varepsilon_{s}=\varepsilon_{d}=10,$ $u_{0}=u_{2}=$
$u_{2}^{\prime}=0,$
$v_{0}=1.5,$ $v_{2}=1.3$}%
\label{fig:Bogo}%
\end{figure}

As shown in Ref.~\cite{Iachello} an analytical solution could be
obtained either for phase $I$ or phase $II$, but in intermediate
situations one could only obtain an approximate or a numerical
solution. Since we have three kinds of bosons in the model and a
macroscopic number of particles, we may use the Bogoliubov
approximation \cite{Bogoliubov}. This approximation consists of
replacing in the Hamiltonian (\ref{ham_iach}) the operators which
create and annihilate bosons in the condensed state by a c-number
and neglecting all terms higher than degree two in operators which
annihilate and create bosons in other states. After this replacement
the Hamiltonian becomes the bilinear
form%
\begin{eqnarray}
H_{d}  &=& H_{0}+\left(  \varepsilon_{s}+v_{0}N_{d}-\mu\right)  \hat{n}%
_{s}+\label{Bogo-d}\\
 &+&\frac{1}{2}v_{2}\sqrt{N_{d}^{2}-l^{2}}\left(  \hat{s}^{\dagger}\hat
{s}^{\dagger}+\hat{s}\hat{s}\right) \nonumber\\
H_{0}  &=& \left(  \varepsilon_{d}-\mu\right)
N_{d}+\frac{1}{4}\left(
2u_{2}+u_{2}^{\prime}\right)  N_{d}^{2}+\frac{1}{4}\left(  2u_{2}%
-u_{2}^{\prime}\right)  l^{2}\nonumber
\end{eqnarray}
where $\hat{d}_{\pm},\hat{d}_{\pm}^{\dag}\rightarrow\sqrt{N_{\pm}}$,
$N_{d}=N_{+}+N_{-},$ $l=N_{+}-N_{-}$ \ and $\mu$ is a chemical potential. The
chemical potential is determined as usual from the condition on the mean total
number of particles. $N_{d}$ is determined by variation. To diagonalize the
bilinear form (\ref{Bogo-d}) we introduce the squeezing transformation%
\[
U_{s}=\exp\frac{\alpha}{2}\left(  \hat{s}^{\dagger2}-\hat{s}^{2}\right)  .
\]
We compare the Bogoliubov approximation with the exact numerical solution in
Fig \ref{fig:Bogo}.%
The right column of Fig~\ref{fig:Bogo} shows the occupancy of the
\emph{s-}bosons. We observe different regions of the parameter corresponding
to different phases of the system. The left column shows that the Bogoliubov
approximation gives remarkably good results either when the system is in phase
$Ia$ or phase $II.$ The failure of the Bogoliubov approximation in phase $Ib$
is expected since the \emph{s-}level is not macroscopically occupied in that
phase.  (Another Bogoliubov approximation based on macroscopic occupancy of only
\emph{d-}bosons could be constructed for phase $Ib$)

In order to investigate the transition between phase $I$ and phase $II$ the
original Hamiltonian (\ref{ham_ns_nd}) can be reduced to the form%
\begin{equation}
H=\varepsilon\hat{n}_{d}-\frac{v_{2}}{2}\hat{W}^{2} \label{Reduced_Ham}%
\end{equation}
by setting $\varepsilon_{s}=v_{0}=2u_{2}=-2u_{2}^{\prime}=-v_{2},$ $u_{0}=0$
and $\varepsilon=\varepsilon_{d}+\frac{1}{2}v_{2}.$ This Hamiltonian contains
the essential features required to describe the transition between the phases.
The system is in phase $I$ for $v_{2}=0$ and it is in phase $II$ for
$\varepsilon=0.$ In the first case the system reduces to a simple two level
system of $N$ non-interacting bosons, and in the second case we have pure
pairing interaction. In order to have a quantitative measure of \emph{s-}wave
and \emph{d-}wave symmetry we define two fractional weight operators
\[
\hat{g}_{d}\equiv\frac{1}{N}\hat{n}_{d}\qquad\hat{g}_{s}\equiv\frac{1}{N}%
\hat{n}_{s},
\]
and the fractional weight measures of a state $\left\vert \Psi\right\rangle $%
\[
S_{\Psi}=\left\langle \Psi\right\vert \hat{g}_{s}\left\vert \Psi\right\rangle
\qquad D_{\Psi}=\left\langle \Psi\right\vert \hat{g}_{d}\left\vert
\Psi\right\rangle
\]
with the properties%
\begin{eqnarray*}
D_{0}  &=&1,\quad D_{N/2}=\frac{1}{2},\quad D_{N}=0\\
S_{0}  &=&0,\quad S_{N/2}=\frac{1}{2},\quad S_{N}=1
\end{eqnarray*}
where the subscripts $0,\frac{N}{2}$ and $N$ indicate $\left\vert
\Psi\right\rangle =\left\vert 0,N\right\rangle ,$ $\left\vert \frac{N}%
{2},\frac{N}{2}\right\rangle ,$ $\left\vert N,0\right\rangle $ respectively.

For pure \emph{s-}wave symmetry $S_{\Psi}$ and $D_{\Psi}$ yield one and zero,
respectively, and vice-versa for pure \emph{d-}wave symmetry. For mixed
symmetries, $S$ and $D$ vary between zero and one.

A phase diagram of the system described by the reduced Hamiltonian
(\ref{Reduced_Ham}) may be obtained readily by calculating the \emph{s-}wave
or the \emph{d-}wave symmetry measure in the exact ground state for different
values of $v_{2}$ and $\varepsilon.$%

\begin{figure}[t]
\includegraphics[
height=6.0275cm,
width=6.0934cm
]%
{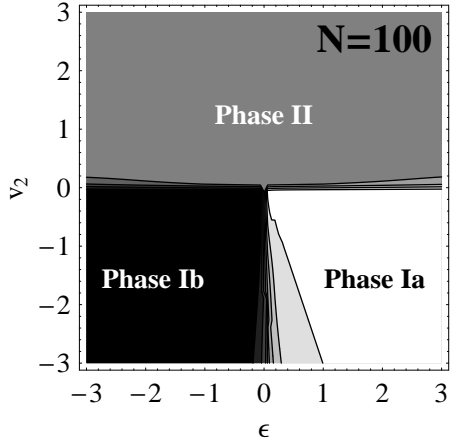}%
\caption{S-wave symmetry measure of the ground state calculated for various
values of $\varepsilon$ and $v_{2}.$ Lighter tones denote higher values.}%
\label{fig:phase}%
\end{figure}
As seen from Fig.~\ref{fig:phase} the system has three distinct
phases. When the system is in phase $I$ the number of \emph{s-} and
\emph{d-} bosons is definite; \ when it is in phase $II$ the number
of \emph{s-} and \emph{d-}bosons is indefinite. The transition
between the phases can be controlled by changing $\varepsilon$ or
$v_{2}.$

The strength of the scattering interaction $\left(  v_{2}\right)  $ could be
controlled by the doping of the superconducting sample. It is known that
doping affects electron-electron and electron-phonon scattering \cite{Yeh},
therefore the desired scattering on the Fermi surface could be achieved by
varying the doping of the sample (Meir Weger, private communication).

\begin{figure}
\begin{minipage}[t]{0.45\textwidth}
\includegraphics[width=0.9\textwidth]{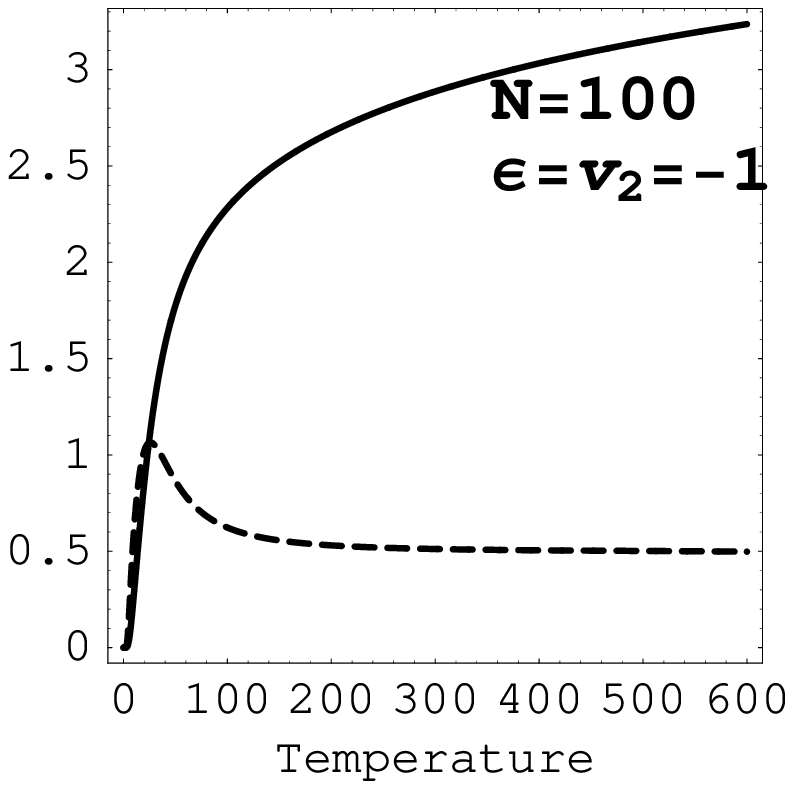}
\caption{The entropy (solid line) and the specific heat of the
system (dashed line) as a function of the temperature. For $T=0$ the
system is in phase $Ib.$} \label{fig:sctemp}
\end{minipage}
\hfill
\begin{minipage}[t]{0.45\textwidth}
\includegraphics[width=\textwidth]{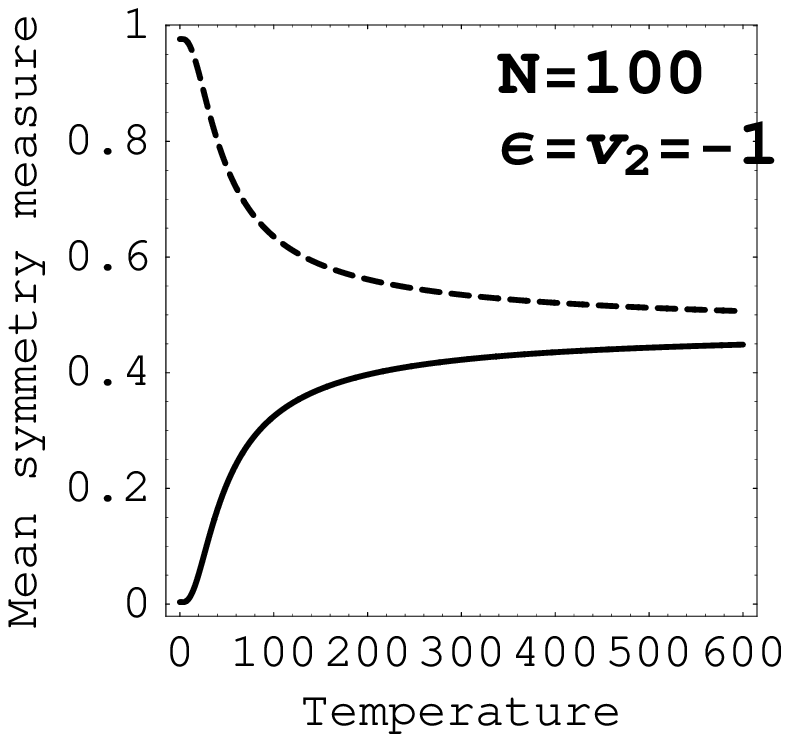}
\caption{Mean symmetry measures as functions of the temperature. The
solid line denotes the s-wave symmetry measure and the dashed line\
denotes the d-wave symmetry measure. For $T=0$ the system is in
phase $Ib.$} \label{fig:symtemp}
\end{minipage}
\hfill
\end{figure}

Having the exact numerical solution we can examine the thermodynamical
properties of this model by calculating the free energy and the density matrix
of the system at inverse temperature $\beta$%
\[
Z=\Tr e^{-\beta H}\qquad F=-\frac{1}{\beta}\ln Z\qquad\hat{\rho
}=\frac{1}{Z}e^{-\beta H}%
\]
We can then explore the behavior of the derivatives of the free energy with
respect to the temperature. The first and the second derivatives of the free
energy are shown to be continuous functions at the explored domain of
parameters. However, there is a jump in the second derivative when the
temperature scale is of the same order as the difference between the ground
and first excited states (Fig \ref{fig:sctemp}).%

This jump does not become a singularity point when the number of
particles tends to infinity, since the width of the peak does not
scale as $N^{-\alpha}$, and therefore there is no second order phase
transition at this point\cite{Pathria}.

Although there is no second order phase transition, changing the temperature
does produce a transition between phases having different symmetries. If the
system is in phase $I$ at $T=0,$ we can calculate the mean ensemble
\emph{s-}wave and \emph{d-}wave symmetry measures as a function of the
temperature%
\[
\bar{S}=\Tr\left(  \hat{g}_{s}\hat{\rho}\right)  \qquad\bar
{D}=\Tr\left(  \hat{g}_{d}\hat{\rho}\right)  .
\]

In Fig.~\ref{fig:symtemp} we present the temperature dependence of $\bar{S}$
and $\bar{D}.$ We observe that both symmetry measures approach $\frac{1}{2}$
when the temperature is increased, which corresponds to phase $II$ of the
system. If the system is initially in phase $II$, it would remain in this
phase for any temperature. It is tempting to propose the temperature as the
control parameter for the transition, but some things should be kept in mind.
First, the temperature scale of the transition may be higher than the critical
temperature of the superconductor. This can only be known after the model is
fit to experimental data and the values of the Hamiltonian parameters
(\ref{Reduced_Ham}) are obtained. Second, if we would like the temperature to
be the control parameter of the transition, there must be a temperature
gradient within the copper-oxide plane of the superconductor, so that the bulk
of the superconductor will have higher temperature than its surface. \ This is
not likely to happen since the superconductor is in thermal equilibrium and
the typical surface depth is of the order of only several atomic layers.

To sum up, in this work we have used a theoretical model proposed by
Iachello \cite{Iachello} for a transition from \emph{d-}wave to
mixed \emph{d-} and \emph{s-}wave symmetry. A phase diagram of the
model was obtained, which showed a distinct separation between the
three phases. Doping of the superconducting sample and temperature
were proposed as possible control parameters for the phases of the
system.

\section*{Acknowledgements} We would like to thank Meir Weger for his
considerable help. JLB thanks the Institute of Theoretical Physics
of Technion for its support and hospitality. Also support from
PSC-CUNY is acknowledged.

\end{document}